\documentclass[prl,twocolumn,floatfix,superscriptaddress,amsmath,citeautoscript,aps,longbibliography]{revtex4-2}
\pdfoutput=1
\usepackage[T1]{fontenc}\usepackage[latin1]{inputenc}
\usepackage{dcolumn,bm,graphicx,color,booktabs,microtype,afterpage} \graphicspath{{Figures/}}
\usepackage[charter,greekuppercase=italicized]{mathdesign}
\usepackage{siunitx}
\renewcommand{\figurename}{Fig.}
\renewcommand{\tablename}{Table}
\makeatletter\renewcommand{\fnum@figure}[1]{\figurename~\thefigure.}\makeatother
\makeatletter\renewcommand{\fnum@table}[1]{\tablename~\thetable.}\makeatother
\newcount\hh \newcount\mm
\hh=\time \divide\hh by 60
\mm=\hh \multiply\mm by 60 \mm=-\mm
\advance\mm by \time
\def\now{\number\hh:\ifnum\mm<10{}0\fi\number\mm}
\usepackage[colorlinks,plainpages=false,linkcolor=blue,urlcolor=blue,citecolor=blue,pdfpagemode=UseNone,pdfstartview=FitBH]{hyperref}

\newcommand{\mnf}{MnF$_2$}
\DeclareSIUnit\angstrom{\text{\fontfamily{cmr}\selectfont{\AA}}} 

\begin{document}

\title{Absence of altermagnetic magnon band splitting in \mnf}

\author{V.~C.~Morano}
\email{vincent.morano@psi.ch}
\affiliation{PSI Center for Neutron and Muon Sciences, Forschungsstrasse 111, 5232 Villigen, PSI, Switzerland}
\author{Z.~Maesen}
\affiliation{PSI Center for Neutron and Muon Sciences, Forschungsstrasse 111, 5232 Villigen, PSI, Switzerland}
\author{S.~E.~Nikitin}
\affiliation{PSI Center for Neutron and Muon Sciences, Forschungsstrasse 111, 5232 Villigen, PSI, Switzerland}
\author{J.~Lass}
\affiliation{PSI Center for Neutron and Muon Sciences, Forschungsstrasse 111, 5232 Villigen, PSI, Switzerland}
\author{D.~G.~Mazzone}
\affiliation{PSI Center for Neutron and Muon Sciences, Forschungsstrasse 111, 5232 Villigen, PSI, Switzerland}
\author{O.~Zaharko}
\affiliation{PSI Center for Neutron and Muon Sciences, Forschungsstrasse 111, 5232 Villigen, PSI, Switzerland}

\date{\today}

\begin{abstract}
Altermagnets are collinear compensated magnets in which the magnetic sublattices are related by rotation rather than translation or inversion. One of the quintessential properties of altermagnets is the presence of split chiral magnon modes. Recently, such modes have been predicted in \mnf. Here, we report inelastic neutron scattering results on an \mnf ~single-crystal along high-symmetry Brillouin zone paths for which the magnon splitting is expected. Within the resolution of our measurement, we do not observe the predicted splitting. The inelastic spectrum is well-modeled using $J_1, ~J_2, ~J_3$ nearest-neighbor exchange interactions with weak uniaxial anisotropy. These interactions have higher symmetry than the crystal lattice, while the interactions predicted to produce the altermagnetic splitting are negligibly small. Therefore, the two magnon modes appear to be degenerate over the entire Brillouin zone and the spin dynamics of MnF$_2$ is indistinguishable from a classical N\'eel antiferromagnet. Application of magnetic field causes a Zeeman splitting of the magnon modes close to the $\mathrm{\Gamma}$ point. Even if chiral magnon modes are allowed by altermagnetic symmetry, the splitting in real materials such as \mnf ~can be negligibly small.
\end{abstract}

\maketitle

\begin{figure*}[ht]
    \centering
    \includegraphics[width=0.95\textwidth]{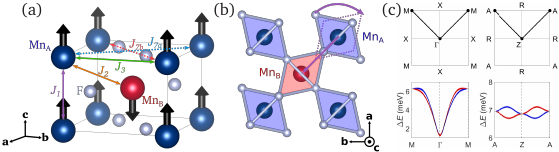}
    \caption{(a)~\mnf ~magnetic structure with opposite spin sublattices labeled as Mn$_{\mathrm{A}}$ and Mn$_{\mathrm{B}}$. The three nearest-neighbor interactions $J_1$, $J_2$ and $J_3$, corresponding to distances of \SI{3.30}{\angstrom}, \SI{3.82}{\angstrom} and \SI{4.87}{\angstrom}, respectively, used in our model Hamiltonian are indicated~\cite{Jauch1988}. The diagonal interactions $J_{7a}$ and $J_{7b}$, corresponding to a distance of \SI{6.89}{\angstrom}, are marked as well.
    (b)~The crystal structure with the $c$-axis oriented perpendicular to the page. The opposite spin sublattices cannot be related by translation or inversion combined with time-reversal, rather the surrounding F ions require 90$^\circ$ rotation in addition to $\left[\frac{1}{2} \frac{1}{2} \frac{1}{2}\right]$ translation. 
    (c)~Simulated dispersion from the $J_{1-3}+D_{\mathrm{c}}$ Hamiltonian, with the addition of low-symmetry interactions $J_{7\mathrm{a}} = -J_{7\mathrm{b}} = \SI{20}{\micro\eV}$ that split chiral magnon modes. The upper two panels indicate the BZ paths traversed in the lower two panels. The chirality of each mode is indicated by red and blue; it alternates upon crossing $\mathrm{\Gamma}$ and $\mathrm{Z}$. The difference between $J_{7a}$ and $J_{7b}$ controls the splitting along the $\overline{\mathrm{\Gamma} \mathrm{M}}$ and $\overline{\mathrm{Z} \mathrm{A}}$ paths.}
    \label{fig:Fig1}
\end{figure*}

Altermagnetism has been proposed as a new classification of collinear magnetism distinct from traditional categories such as ferromagnetism, antiferromagnetism, and ferrimagnetism~\cite{Smejkal2022a, Smejkal2022b, Smejkal2022c, Mazin2021, Mazin2024}. While various definitions have been put forward, generally altermagnets are understood as materials that break Kramers spin degeneracy yielding spin-polarized band structures despite possessing compensated magnetic order with no net magnetization. Time-reversal symmetry breaking in altermagnets is a consequence of the opposite-spin altermagnetic sublattices being related by crystal symmetries such as rotation rather than translation or inversion. Much like the electronic excitations, chiral magnon modes are predicted to split beyond particular high-symmetry planes of the Brillouin zone (BZ) even in the absence of spin-orbit coupling~\cite{Smejkal2022b}. The chirality of magnon modes alternates in reciprocal space realizing a magnetic analog of superconductors with different chirality \cite{mazin2022notes}. These modes are of both fundamental and technological interest, potentially contributing to the development of high-frequency spintronics devices. In principle, chiral magnon modes are directly observable by polarized neutron scattering~\cite{mcclarty2024observing}. The magnitude of the chiral magnon splitting, however, depends on the relative magnitudes of exchange interactions that break the degeneracy of the chiral modes in the BZ.

Recent inelastic neutron scattering (INS) results support the existence of chiral magnon modes in $\alpha$-MnTe~\cite{liu2024chiral}. On the other hand, studies of another altermagnetic candidate, rutile RuO$_2$, have cast its own (alter)magnetic status into doubt~\cite{Hiraishi2024, Huang2024, Liu2024}. \mnf ~is a rutile compound that has been studied with neutron scattering for over seven decades~\cite{Shull1949, Ruderman1949, Shull1951, Erickson1953, Alperin1962, okazaki1964neutron, turberfield1965development, Nikotin1969, schweika2002longitudinal, Yamani2010}. Traditionally, it has been understood as a N\'eel antiferromaget in which the ($S = 5/2$, $L = 0$) Mn$^{2+}$ spins order along the $c$-axis \cite{Erickson1953} below $T_{\mathrm{N}} = 67$~K \cite{Heller1962}. The observed magnetic excitations are well-described by a $J_1$, $J_2$, $J_3$ model with weak $c$-axis anisotropy, which will be called the $J_{1-3}+D_{\mathrm{c}}$ model. This is given by the following Hamiltonian:

\begin{equation} \label{eq:ham}
    \mathcal{H} = \sum_{ij} J_{ij} \mathbf{S}_i \cdot \mathbf{S}_j + D_{\mathrm{c}} \sum_{i} S_{i,z}^2.
\end{equation}
Here the first sum is taken over only the first three nearest neighbors and $D_{\mathrm{c}}$ is the uniaxial single-ion anisotropy that can be largely understood from the dipole-dipole interaction \cite{Shirane2002}. Despite the simplicity of its magnetic order, \mnf ~displays peculiar domain physics and is notable for being among the first two materials in which piezomagnetism was experimentally observed~\cite{borovik1960}. Recently, extensive theoretical work has reclassified \mnf ~as an altermagnet with split chiral magnon excitations~\cite{Smejkal2022a, Smejkal2022b}. The altermagnetic nature of the modes is not captured by the $J_{1-3}+D_{\mathrm{c}}$ model. Rather, lower-symmetry interactions (referred to as "anisotropic exchange" in e.g. \cite{Smejkal2022b}) beginning at the seventh-nearest-neighbor are required to break the "spurious" sublattice symmetry obeyed by the shorter-range interactions and split the magnon modes \cite{gohlke2023spurious}. While long-range interactions are expected to be weak for this insulating system, sets of exchange couplings on the order of tens of \SI{}{\micro\eV} can potentially split the magnon bands by hundreds of \SI{}{\micro\eV} and could be resolvable by experiment, see Fig.~\ref{fig:Fig1}(c).

\begin{figure*}[ht]
    \centering
    \includegraphics[width=1.0\textwidth]{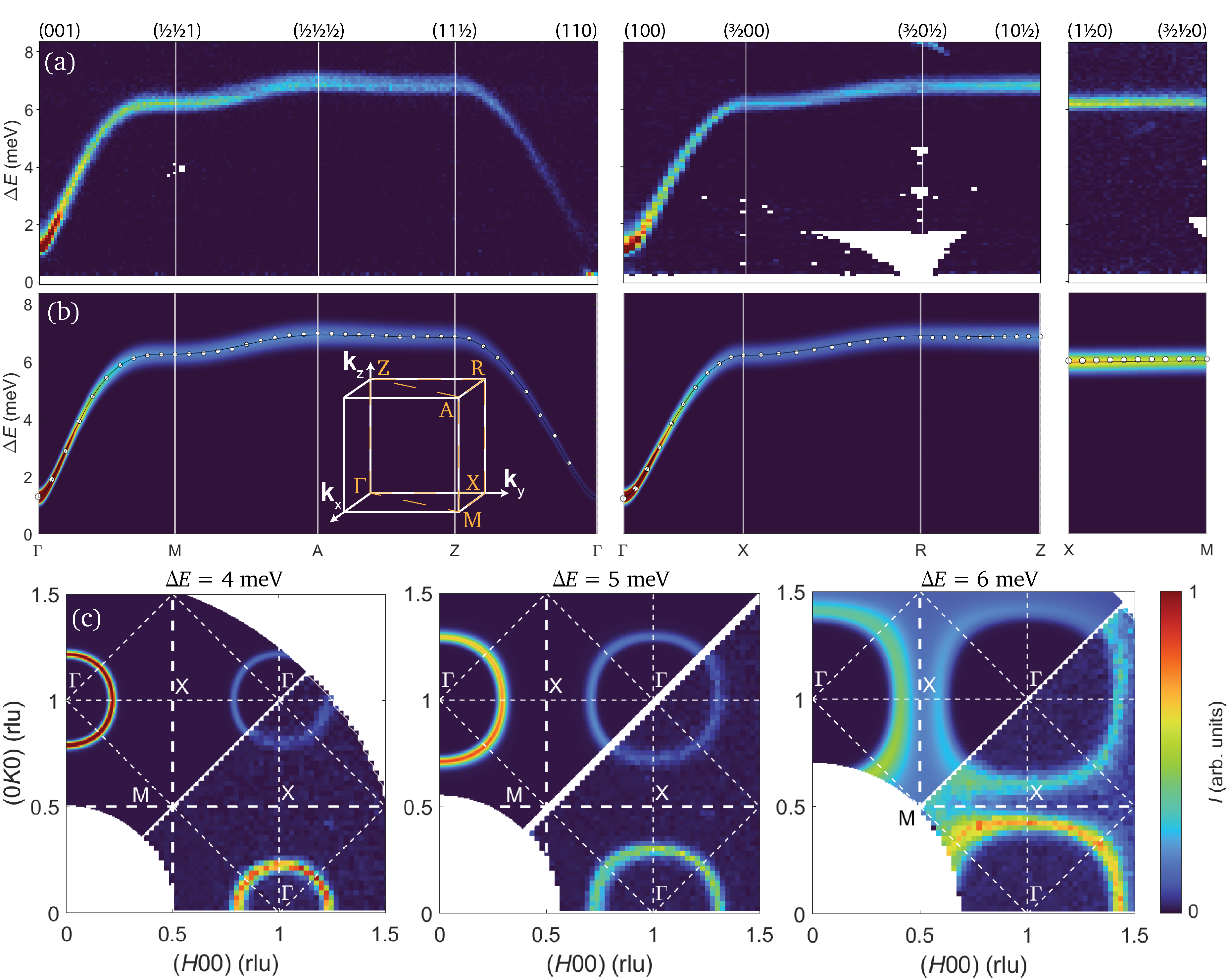}
    \caption{(a)~\mnf ~spin wave spectrum measured along high-symmetry directions as indicated in the figure. The left panel contains $(HHL)$, the middle panel $(H0L)$, and the right panel $(HK0)$ planes. All high-symmetry BZ paths other than $\overline{\mathrm{A} \mathrm{R}}$ are measured. White pixels correspond to datapoints that were not measured. The feature at e.g. 8~meV in $\mathrm{\Delta} E$ and $\mathrm{R}$ in $\mathbf{Q}$ is a Currat-Axe spurion. The data are integrated over 0.1 meV steps along $\mathrm{\Delta} E$ and \SI{0.04}{\per\angstrom} in $\mathbf{Q}$ perpendicular to the plotted axis. The $(HHL)$ and $(HK0)$ data are integrated over \SI{0.02}{\per\angstrom} steps in $\mathbf{Q}$ along the plotted axis while the $(H0L)$ data are integrated over \SI{0.03}{\per\angstrom} steps in $\mathbf{Q}$ along the plotted axis. Each panel, from different scattering planes, has a different overall scale factor applied to the intensity.
    (b)~Simulated spin wave spectrum using the best fit Hamiltonian and notation of the points in the Brillouin zone. The size of each point corresponds to its observed amplitude when fit to a Gaussian peak. Each panel, from different scattering planes, has a different overall scale factor applied to the intensity.
    (c)~Constant-$\mathrm{\Delta} E$ slices in $(HK0)$ showing conical excitations dispersing from $\mathrm{\Gamma}$ that become anisotropic upon merging at higher $\mathrm{\Delta} E$. The bottom-right part of each panel shows INS data and top-left demonstrates calculated intensities. The data were integrated over \SI{0.1}{\milli\eV} and \SI{0.03}{\per\angstrom} along $(H00)$ and $(0K0)$. The colorbar applies to all panels in (a), (b) and (c).}
    \label{fig:Fig2}
\end{figure*}

\begin{figure}[ht]
    \centering
    \includegraphics[width=0.45\textwidth]{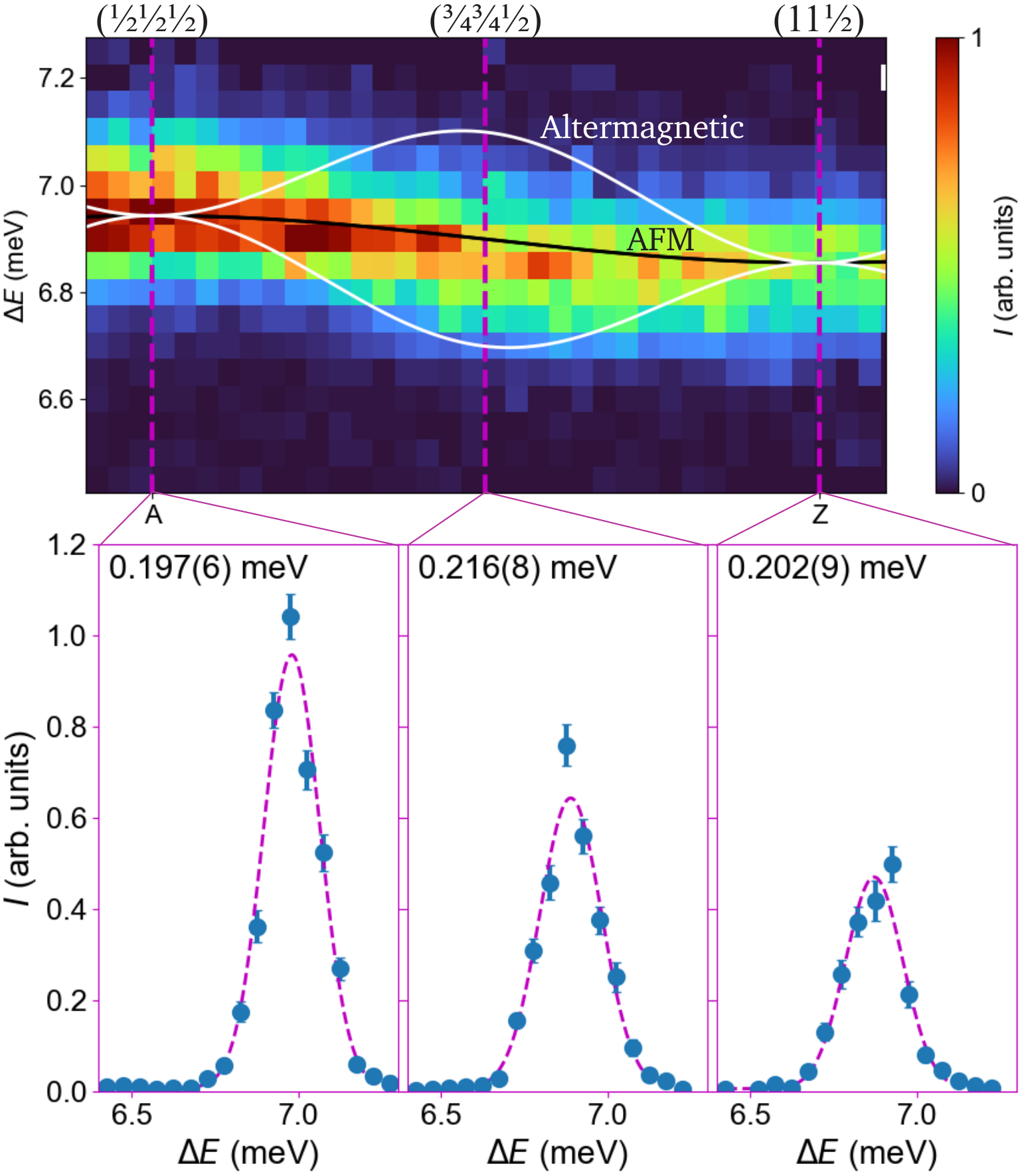}
    \caption{~Slice extending along $\overline{\mathrm{A} \mathrm{Z}}$. For these scans, the slits and incident neutron energy were optimized to increase $\mathrm{\Delta} E$-resolution. The slice is integrated in $\mathrm{\Delta} E$ over \SI{0.05}{\milli\eV} steps and in $\mathbf{Q}$ over \SI{0.03}{\per\angstrom} steps both along and perpendicular to the plotted axis. The dispersion overplotted in black is for the best-fit antiferromagnetic model of Figure \ref{fig:Fig2}(b) while the dispersion in white is the altermagnetic Hamiltonian of Figure \ref{fig:Fig1}(c). $\mathrm{\Delta} E$ cuts in magenta are shown at $\left( \frac{3}{4} \frac{3}{4} \frac{1}{2} \right)$, the midpoint along $\overline{\mathrm{A} \mathrm{Z}}$ at which the splitting is expected, as well as $\mathrm{A}$ and $\mathrm{Z}$, where no splitting is expected. Each panel lists the best-fit Gaussian FWHM in $\mathrm{\Delta} E$, each cut was binned with 0.05 meV steps along $\mathrm{\Delta} E$ and \SI{0.025}{\per\angstrom} width in the perpendicular directions. There is no statistically significant difference ($p > 0.05$) between the $\mathrm{\Delta} E$-widths, which are resolution-limited.}
    \label{fig:Fig3}
\end{figure}

Here we report unpolarized inelastic neutron scattering results on \mnf ~intended to quantify the altermagnetic splitting. We provide the most complete INS dataset on the spin dynamics in MnF$_2$ measured in three scattering planes. We fit the exchange Hamiltonian to the observed spin wave excitations using linear spin-wave theory (LSWT) and examine high-resolution scans through BZ points for which splitting is expected. The altermagnetic splitting is not observed, rather the spectrum is degenerate up to the $\mathrm{\Delta} E$-resolution of our experiment. Finally, we report changes to the magnon band structure under application of a 10~T magnetic field along $(H \overline{H} 0)$ in reciprocal lattice units (rlu) which comply with the $J_{1-3}+D_{\mathrm{c}}$ model.

A $m = \SI{3.446(1)}{\gram}$ \mnf ~single-crystal was aligned sequentially in the $(HHL)$, $(HK0)$ and $(H0L)$ horizontal scattering planes. For the measurements in $(HHL)$ and $(HK0)$, the sample was loaded into an 11~T vertical-field cryomagnet. For the measurement in $(H0L)$, the sample was loaded into an Orange ILL cryostat. All experiments were performed on the CAMEA multiplexing spectrometer at the Paul Scherrer Institut~\cite{CAMEA}. See the Supplemental Material (SM) for details on the scans. Data reduction and resolution calculations were performed in Python \cite{anaconda} with the MJOLNIR software package~\cite{lass2020mjolnir}. LSWT calculations and fitting were performed in MATLAB~\cite{matlab} with the SpinW software package~\cite{toth15}. All data were symmetrized by folding into the positive $(HKL)$ octant. Data and code are publicly available at \cite{github}.

Rutile \mnf ~adopts a tetragonal structure (space group $P4_2/mnm$ or 136 \cite{Jauch1988}) in which the Mn ions at the 2a Wyckoff position are related by a 4-fold rotation about the $c$-axis followed by a $\left[ \frac{1}{2} \frac{1}{2} \frac{1}{2} \right]$ translation [Fig.~\ref{fig:Fig1}(a,b)]. \mnf ~is therefore altermagnetic when it undergoes collinear ordering below $T = 67$~K. Consistent with this classification, exchange interactions start at the seventh nearest-neighbor splitting the magnon spectrum along paths connecting $\mathrm{\Gamma}$ to $\mathrm{M}$ ($\overline{\mathrm{\Gamma} \mathrm{M}}$) and $\mathrm{Z}$ to $\mathrm{A}$ ($\overline{\mathrm{Z} \mathrm{A}}$) [Fig.~\ref{fig:Fig1}(c)]. Beyond the bilinear exchange interactions, magnetoelastic contributions modeled as biquadratic exchange terms in the Hamiltonian can also contribute to the splitting, though these are not considered in this work. If each pair of bilinear interactions has a magnitude of tens of \SI{}{\micro\eV} and opposite relative sign, the splitting is expected to be observable with inelastic neutron scattering on CAMEA which has a resolution on the order of \SI{400}{\micro\eV}. If the magnitude and/or relative signs of the interactions are unfavorable and the chiral magnon splitting is small relative to the $\mathrm{\Delta} E$-resolution available, however, the higher-symmetry nearest-neighbor interactions dominate. The spectrum will appear to have a "spurious," or approximate, degeneracy \cite{gohlke2023spurious, Smejkal2023} and the spin dynamics will be experimentally indistinguishable from a classical N\'eel antiferromagnet.

Figure~\ref{fig:Fig2}(a) shows the experimental dispersion obtained at 1.5~K and 0~T along high-symmetry paths through the Brillouin zone. Consistent with earlier reports ~\cite{okazaki1964neutron, Nikotin1969, Shirane2002, Yamani2010, CAMEA}, we find a single gapped spin wave mode extending from $\mathrm{\Delta} E = 1$ to 7~meV energy transfer. Authors of~\cite{Nikotin1969} have predicted a lifting of the magnon degeneracy along particular BZ paths due to dipole-dipole interactions, but were unable to resolve it experimentally. While the degeneracy along $\overline{\mathrm{\Gamma} \mathrm{M}}$ may also be lifted by this anisotropy, complicating experimental quantification of altermagnetic splitting, the degeneracy along $\overline{\mathrm{Z} \mathrm{A}}$ is unaffected. The dispersion observed along $\overline{\mathrm{Z} \mathrm{A}}$ (Fig. \ref{fig:Fig3}) is inconsistent with a $J_1$, $J_2$ model and implies a ferromagnetic (FM) $J_3$ interaction~\cite{Tonegawa1969, Nikotin1969} (in contrast to the antiferromagnetic (AFM) $J_3$ reported in~\cite{Nikotin1969}). We therefore fit the observed dispersion to an effective $J_{1-3}+D_{\mathrm{c}}$ Hamiltonian (for details see SM). The resulting FM $J_1 = \SI{-67.7(9)}{\micro\eV}$, AFM $J_2 = \SI{302.2(6)}{\micro\eV}$, FM $J_3 = \SI{-4.4(4)}{\micro\eV}$ and easy-axis $D_{\mathrm{c}} = \SI{-26.7(6)}{\micro\eV}$ reproduce the observed dispersion. The $J$ parameters, modulo a factor of 2 depending on the definition of the Hamiltonian, are broadly consistent with earlier reports~\cite{okazaki1964neutron, Nikotin1969, Shirane2002, Yamani2010, CAMEA}. Figure~\ref{fig:Fig2}(b) is the inelastic spectrum simulated from these best-fit parameters. The white points indicate experimental $\mathrm{\Delta} E$ values obtained from cuts at a given $\mathbf{Q}$, which the model was fit to. Although the magnitude of the exchange interactions is not solely dependent on distance, we note that the magnitude of $J_3$ at \SI{4.87}{\angstrom} has already fallen to a few \SI{}{\micro\eV}, while $J_7$ lies another 2~\AA\ farther away (\SI{6.89}{\angstrom}) and is therefore expected to be small. Figure~\ref{fig:Fig2}(c) shows constant-$\mathrm{\Delta} E$ slices through the modes dispersing from the $\mathrm{\Gamma}$-point, with the corresponding fits displayed in the same frame demonstrating the good quantitative agreement between our data and the model.

As the next step, we quantify the altermagnetic magnon splitting in \mnf. Figure~\ref{fig:Fig2}(a) shows the spin wave spectra obtained along the $\overline{\mathrm{\Gamma} \mathrm{M}}$ and $\overline{\mathrm{Z} \mathrm{A}}$ paths where splitting has been calculated. We observe a single mode extending along both paths. Even in the absence of clear splitting, broadening of the modes along such paths could indicate an energy difference between the two modes that is small relative to the $\mathrm{\Delta} E$-resolution of the measurement. To quantify the broadening, the $\overline{\mathrm{Z} \mathrm{A}}$ path was measured after optimizing the configuration of CAMEA for the best resolution at $\mathrm{\Delta} E = 6.9$~meV by narrowing the slits at the virtual source from \SI{40}{\milli\meter} to \SI{10}{\milli\meter}, narrowing the vertical sample slits from \SI{32}{\milli\meter} to \SI{10}{\milli\meter}, and adjusting $E_{\mathrm{i}}$ so that the excitations are measured on the  $E_{\mathrm{f}} = 3.5$~meV rather than the $E_{\mathrm{f}} = 4.6$~meV analyzer. The elastic line signal width improved from about $\mathrm{FWHM} = \SI{160}{\micro\eV}$ to $\mathrm{FWHM} = \SI{90}{\micro\eV}$. Figure~\ref{fig:Fig3} shows a $\mathrm{\Delta} E$ cut at $\mathbf{Q} = \left( \frac{3}{4} \frac{3}{4} \frac{1}{2} \right)$ where the altermagnetic splitting is expected to be maximal, and compares it to cuts at $\mathbf{Q} = \left(1 1 \frac{1}{2} \right)$ and $\left( \frac{1}{2} \frac{1}{2} \frac{1}{2} \right)$, where the modes would be degenerate even in the presence of the altermagnetic coupling. There is no statistically significant difference between the widths. The $\mathrm{FWHM}$ are narrow relative to the estimated signal width at $\mathrm{\Delta} E = 6.9$~meV, suggesting the modes are resolution-limited. For the $J_{1-3}+D_{\mathrm{c}}$ model discussed above with the addition of altermagnetic $J_7$ interactions having opposite sign, the absence of broadening at the level of $\mathrm{FWHM} = 0.25$~meV implies $|J_{7\mathrm{a}} - J_{7\mathrm{b}}| < \SI{12}{\micro\eV}$ and a splitting of the chiral modes by $<\SI{120}{\micro\eV}$. For simplicity we neglect longer-range altermagnetic interactions like $J_8$, though in principle they may also contribute to the splitting.

\begin{figure}[ht]
    \centering
    \includegraphics[width=0.48\textwidth]{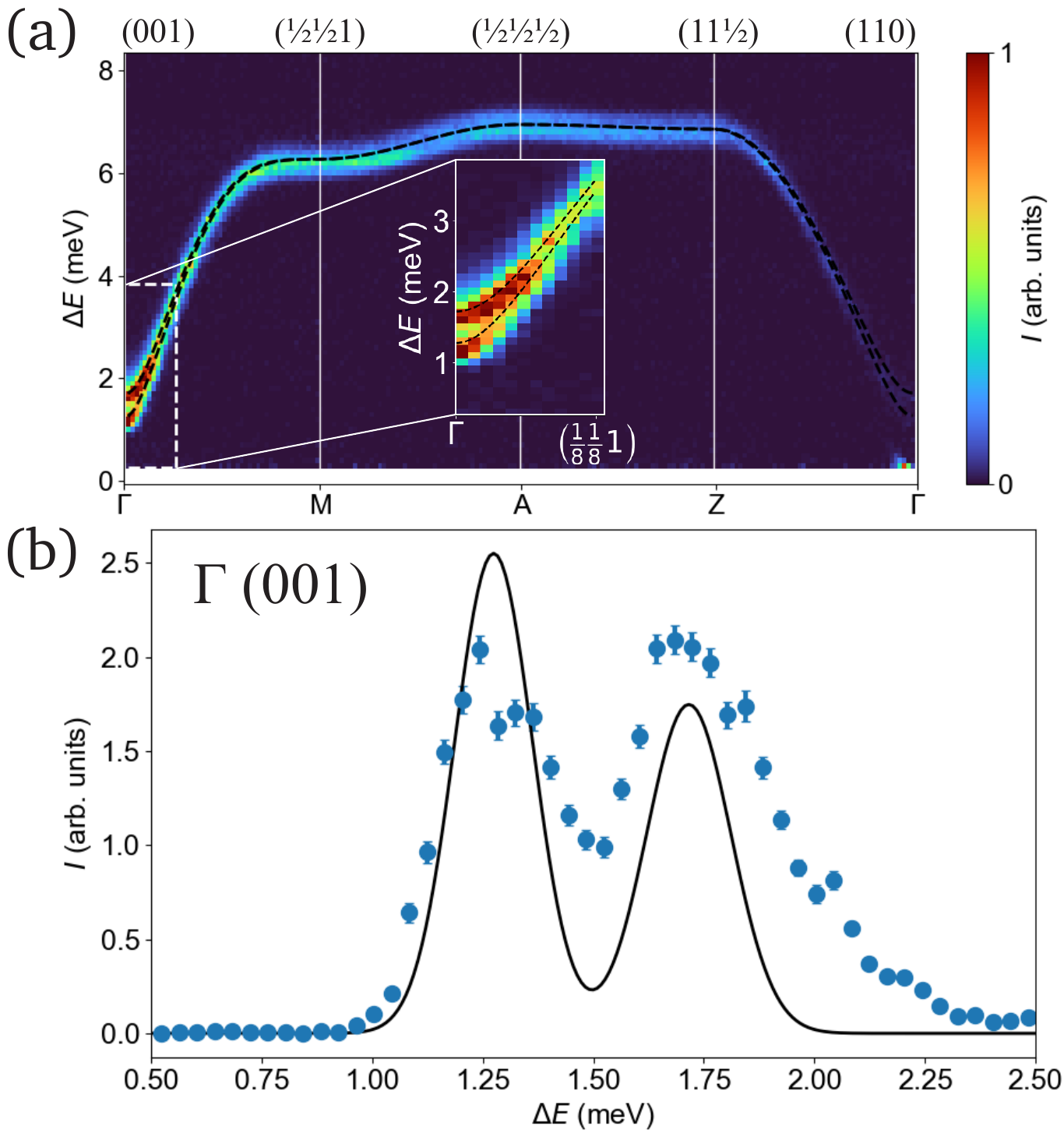}
    \caption{(a)~Spectrum along high-symmetry directions obtained in the $(HHL)$ scattering plane at 10~T. The overplotted dispersion was obtained from the best fit parameters described in Fig.~\ref{fig:Fig2} after relaxing the magnetic structure in a 10~T field applied along $(H\overline{H}0)$. The data are integrated over 0.1 meV steps along $\mathrm{\Delta} E$, \SI{0.04}{\per\angstrom} in $\mathbf{Q}$ perpendicular to the plotted axis and \SI{0.02}{\per\angstrom} in $\mathbf{Q}$ along the plotted axis. The inset focuses on the splitting at $\mathrm{\Gamma}$.
    (b)~$\mathrm{\Delta} E$ cut examining the splitting at $\mathrm{\Gamma}$ with the data integrated over 0.04 meV steps along $\mathrm{\Delta} E$ and \SI{0.05}{\per\angstrom} steps in the perpendicular directions. The solid black line is a $\mathrm{\Delta} E$ cut through the simulated spectrum convolved in $\mathrm{\Delta} E$ with a $\mathrm{FWHM} = \SI{200}{\micro\eV}$ Gaussian, but with no convolution or binning in $\mathbf{Q}$.}
    \label{fig:Fig4}
\end{figure}

Here we comment on two other sources, beyond altermagnetic interactions, that can split the otherwise degenerate magnon modes. First, while the maximum splitting in the $(H0L)$ scattering plane due to dipole-dipole interactions is anticipated to be on the order of \SI{5}{\micro\eV} and was not observed by inelastic neutron scattering \cite{Nikotin1969} in that plane, we estimate the splitting at $\mathrm{M}$ to be about \SI{180}{\micro\eV}. While our high-resolution scans are unable to definitively resolve the splitting, constant-$\mathbf{Q}$ cuts along the $\mathrm{A}$ (no dipole-dipole splitting expected) to $\mathrm{M}$ (dipole-dipole splitting expected) path show a broadening of the mode by about \SI{100}{\micro\eV}. Second, application of an external magnetic field may modify the exchange interactions and/or the symmetry via the magnetoelastic coupling and split degenerate modes. Fig.~\ref{fig:Fig4}(a) shows the magnetic excitation spectrum obtained with a 10~T field applied along $(H\overline{H}0)$. We find the spectrum is split at $\mathrm{\Gamma}$ [Fig.~\ref{fig:Fig4}(b)], with a lower and upper \cite{Hagiwara1996} mode emerging as expected for a canted antiferromagnet. We note that this splitting is distinct from the altermagnetic splitting that has been the focus of recent theory work. The spectrum in field can be accurately described using the $J_{1-3}+D_{\mathrm{c}}$ model and $g = 2$ \cite{Hagiwara1996}, with the magnetic structure relaxed after applying the 10~T field to result in the canted state. No splitting of the altermagnetic modes along $\overline{\mathrm{Z} \mathrm{A}}$ under field is observed, which means that despite the presence of magnetoelastic coupling~\cite{borovik1960} application of 10~T field along $(H\overline{H}0)$ does not change the exchange interactions beyond our resolution. The resulting dispersion is plotted on top of the data in Fig. \ref{fig:Fig4}, reproducing the splitting at $\mathrm{\Gamma}$.

To conclude, we report inelastic neutron scattering results of \mnf~where chiral magnon modes have been predicted \cite{gohlke2023spurious, mcclarty2024observing}. Up to the highest $\mathrm{\Delta} E$-resolution at $\mathrm{\Delta} E = 6.9$ meV measured on CAMEA, we observe no signs of the proposed splitting. Instead, the modes are "spuriously" degenerate, consistent with an exchange scheme dominated by the first three nearest-neighbor interactions that have higher symmetry than the crystal lattice~\cite{gohlke2023spurious}. We note that our results do not contradict the predictions of Ref. \cite{mcclarty2024observing}. Rather the lower-symmetry, longer-range exchange interactions that are expected to split the magnon modes are weak, have comparable magnitudes, and/or have identical signs. While future experiments at higher resolution may be able to resolve such modes in \mnf ~and the effects of strain on the splitting remain to be investigated, our work underlines the importance of identifying candidate materials in which the scale of the predicted phenomena are accessible to experiment and, ultimately, application.

\textit{Acknowledgments}---
We acknowledge stimulating discussions with Prof. Christof Niedermayer and Dr. Urs Staub. \smallskip \\ The neutron scattering experiments were performed on CAMEA at the Swiss spallation neutron source SINQ, Paul Scherrer Institute, Villigen, Switzerland. The CAMEA proposal number was 20240628. This work was supported by the Swiss National Science Foundation (Grant No. 200021-219950).  

\bibliography{Main}

\end{document}


\title{Supplemental Material for: Absence of altermagnetic magnon band splitting in \mnf}

\author{V.~C.~Morano}
\email{vincent.morano@psi.ch}
\affiliation{PSI Center for Neutron and Muon Sciences, Forschungsstrasse 111, 5232 Villigen, PSI, Switzerland}
\author{Z.~Maesen}
\affiliation{PSI Center for Neutron and Muon Sciences, Forschungsstrasse 111, 5232 Villigen, PSI, Switzerland}
\author{S.~E.~Nikitin}
\affiliation{PSI Center for Neutron and Muon Sciences, Forschungsstrasse 111, 5232 Villigen, PSI, Switzerland}
\author{J.~Lass}
\affiliation{PSI Center for Neutron and Muon Sciences, Forschungsstrasse 111, 5232 Villigen, PSI, Switzerland}
\author{D.~G.~Mazzone}
\affiliation{PSI Center for Neutron and Muon Sciences, Forschungsstrasse 111, 5232 Villigen, PSI, Switzerland}
\author{O.~Zaharko}
\affiliation{PSI Center for Neutron and Muon Sciences, Forschungsstrasse 111, 5232 Villigen, PSI, Switzerland}

\date{\today}

\maketitle

\textit{Inelastic Neutron Scattering}---
180$^{\circ}$ scans with 1$^{\circ}$ steps were measured at 1.6~K and 0~T in the $(HHL)$ and $(HK0)$ orientations. Each rotation angle was counted over a monitor of 50k counts, corresponding to counting times of \SI{11}{\second} at $E_{\mathrm{i}} = 5.5$ meV and 17 s at $E_{\mathrm{i}} = 11.5$ meV. An identical scan was performed in $(HHL)$ with a 10~T field applied along $(H\overline{H}0)$. In the $(H0L)$ orientation, a 100$^{\circ}$ scan with 2$^{\circ}$ steps counted over a monitor of 200k counts was obtained at 1.9~K and 0~T. This monitor corresponds to a counting time of \SI{49}{\second} at $E_{\mathrm{i}} = 5.5$ meV and \SI{76}{\second} at $E_{\mathrm{i}} = 11.5$ meV. These $(H0L)$ scans were previously reported in \cite{CAMEA}. The elastic line full-width at half-maximum ($\mathrm{FWHM}$) on the $E_{\mathrm{f}} = 4.6$~meV analyzer, which measures the $\mathrm{\Delta} E = 6.9$~meV mode where splitting has been predicted [Fig. 1(c)], is about \SI{157}{\micro\eV} (see Table III of \cite{CAMEA}).

Additional high $\mathrm{\Delta} E$-resolution scans probing for the expected splitting at $\mathrm{\Delta} E = 6.9$~meV were performed in the $(HHL)$ scattering plane with the virtual source slits and vertical sample slits narrowed. These high $\mathrm{\Delta} E$-resolution scans were taken over 62$^{\circ}$ with 1$^{\circ}$ steps. The monitor was 75k counts, corresponding to \SI{71}{\second} at $E_{\mathrm{i}} = 10.3$ meV. The elastic line $\mathrm{FWHM}$ on the $E_{\mathrm{f}} = 3.5$~meV analyzer, which measures the $\mathrm{\Delta} E = 6.9$~meV mode, is about \SI{87}{\micro\eV}. The uncertainty in the fitted $\mathrm{FWHM}$ of Figure 3 are the standard errors that increase the best-fit chi-squared $\chi^2$ by the reduced chi-squared $\chi^2_{\nu}$, where $\nu$ is the statistical degrees of freedom \cite{lmfit}.

Altermagnetic splitting in \mnf~can in principle be measured along either $\overline{\mathrm{\Gamma} \mathrm{M}}$ or $\overline{\mathrm{Z} \mathrm{A}}$. On the one hand, the mode along $\overline{\mathrm{\Gamma} \mathrm{M}}$ is located at lower $\mathrm{\Delta} E$ compared to $\overline{\mathrm{Z} \mathrm{A}}$ and can therefore be measured at higher instrumental resolution. On the other hand, $\mathrm{\Delta} E$ varies greatly along $\overline{\mathrm{\Gamma} \mathrm{M}}$. As a result, the $\mathrm{\Delta} E$-dependent resolution ellipsoid is also expected to vary along this path. This change can be especially significant where the mode crosses analyzers. However, along $\overline{\mathrm{Z} \mathrm{A}}$, $\mathrm{\Delta} E$ only changes by about \SI{0.1}{\milli\eV}. To minimize changes in the FWHM (Figure 3) due to resolution effects, we therefore focus on $\overline{\mathrm{Z} \mathrm{A}}$ when quantifying the altermagnetic broadening. This path has the additional advantage of not being affected by the dipole-dipole induced splitting.

\textit{Fitting the Hamiltonian}---
The dispersion is extracted by taking cuts along $\mathrm{\Delta} E$ at constant $\mathbf{Q}$ and fitting the observed mode to a Gaussian. Fits are performed to intensities weighted by a factor of $1/\sigma$. The standard deviation $\sigma$ is calculated under the Gaussian approximation as $\sigma = \sqrt{N}$, where $N$ is the number of counts. Bins with 0 counts are ignored. Each cut with a step size of \SI{0.05}{\milli\eV} along $\mathrm{\Delta} E$ is integrated over \SI{0.03}{\per\angstrom} in $\mathbf{Q}$. The uncertainty in the fitted values $\mathrm{\Delta} E$ is determined from the square root of the corresponding diagonal element of the covariance matrix \cite{scipy}.

We fit the spectrum to an effective $J_{1-3}+D_{\mathrm{c}}$ Hamiltonian. All four parameters, $J_1$, $J_2$, $J_3$ and $D_{\mathrm{c}}$, are fit to 142 BZ points along high-symmetry directions across all three horizontal scattering planes. Note that these fits were performed in SpinW \cite{toth15} rather than by fitting directly to analytical expressions as was done in~\cite{okazaki1964neutron, Shirane2002, Yamani2010, CAMEA}. Uncertainties in the fitted parameters are obtained by iterating the parameter under consideration, fitting the remaining parameters, and taking half the difference between upper and lower threshold values of the iterated parameter at which the best-fit chi-squared $\chi^2$ is increased by the reduced chi-squared $\chi^2_{\nu}$.

Fits are also obtained to a model in which the dipole-dipole interactions are included \cite{Nikotin1969} rather than an effective single-ion anisotropy $D_{\mathrm{c}}$. The goodness-of-fit is slightly worse, so we report the single-ion fits in accordance with most inelastic studies on \mnf~\cite{okazaki1964neutron, Shirane2002, Yamani2010, CAMEA}. Fits are also obtained using magnetic resonance results \cite{Johnson1959, Hagiwara1996} for the gap at $\mathrm{\Gamma}$, where resolution effects are expected to be significant (c.f. the high-$\mathrm{\Delta} E$ tail in Figure 4(b)). Again there is no improvement to the fit, we therefore report fits that include only our own neutron scattering results from the scans described herein.

\bibliography{Supplement}